\documentclass[aps,pra,reprint,showpacs]{revtex4-1}
\usepackage[dvipdfmx]{graphicx}
\usepackage[dvipdfmx]{color}
\usepackage{amsmath,amssymb}
\usepackage{bm}

\newcommand{\cket}[1]{\left|#1\right\rangle}
\newcommand{\bra}[1]{\left\langle#1\right|}
\newcommand{\bracket}[2]{\left\langle#1|#2\right\rangle}

\begin{document}


\title{Upper bound of one-magnon excitation and lower bound of effective mass for ferromagnetic spinor Bose and Fermi gases}


\author{Masaya Kunimi}
\email{E-mail address: kunimi@hs.pc.uec.ac.jp}
\author{Hiroki Saito}
\affiliation{Department of Engineering Science, University of Electro-Communications, Tokyo 182-8585, Japan}


\date{\today}

\begin{abstract}
Using a variational method, we derive an exact upper bound for one-magnon excitation energy in ferromagnetic spinor gases, which limits the quantum corrections to the effective mass of a magnon to be positive. We also derive an upper bound for one-magnon excitation energy in lattice systems. The results hold for both Bose and Fermi systems in $d$ dimensions as long as the interaction is local and invariant under spin rotation.
\end{abstract}

\maketitle

\section{Introduction}\label{sec:Introduction}
Effective mass is one of the important concepts in condensed-matter physics. This appears in various systems, such as particles in a periodic potential \cite{Ashcroft1976}, Fermi liquid \cite{Nozieres1966}, and polaron problems \cite{Feynman1972}. Since the effective mass reflects not only the effects of static structures of systems but also dynamical processes between particles, it is a key quantity to understand the many-body physics. 

Here, we focus on the problem of whether the interactions between particles increase or decrease the effective mass compared with the mass of a bare particle. Liquid ${}^3$He \cite{Nozieres1966} and $f$-electron systems \cite{Hewson1997} are examples that the effective mass is heavier than the bare mass. Conversely, the effective mass is lighter than the bare mass in a high-density electron gas \cite{Gell-Mann1957,Nozieres1966} and a polaron system \cite{Marchand2013}. Thus, it is not obvious that the effective mass is increased or decreased by quantum many-body effects.

Ultracold atomic gases are suitable systems for exploring many-body physics owing to their high experimental controllability \cite{Bloch2008}. In particular, ultracold atomic gases with spin-degrees of freedom have intriguing features originating from superfluidity and magnetism, which have been extensively explored \cite{Kawaguchi2012,Stamper-Kurn2013} since the experimental realization of spinor Bose-Einstein condensates (BECs) \cite{Stamper-Kurn1998,Stenger1998,Barrett2001,Chang2004,Schmaljohann2004,Kuwamoto2004,Pasquiou2011}. In the recent experiment performed by the Berkeley group \cite{Marti2014}, the magnon (spin-wave) dispersion and its effective mass in the spin-1 ferromagnetic BEC were precisely measured. The technique of high-precision measurement opens up the opportunity to study the spin-wave excitations in the spinor gases. According to the mean-field prediction for the magnon excitation, its dispersion is the same as that of the free particle \cite{Ohmi1998,Ho1998}. This indicates that the effective mass of the magnon is the same as the atomic mass at the mean-field level. However, the observed effective mass of the magnon is heavier than that of the atomic mass \cite{Marti2014}. 

In this paper, using a simple variational method, we derive a restriction on the effective mass of a magnon in the ferromagnetic spinor gases: the effective mass is never decreased by the quantum many-body corrections as long as the interactions between particles are local and invariant under spin rotation. We also derive the upper bound for the magnon dispersion relation in a lattice system. These results hold for both Bose and Fermi systems in $d$ dimensions. 

\section{Continuous Systems}\label{eq:sec_continuous_systems}
We consider spin-$f$ bosons or fermions in a $d$-dimensional space. The field operator is denoted by $\hat{\psi}_m(\bm{r})$, where $m=-f, -f+1,\cdots, f$ represents a magnetic sublevel. The field operators satisfy the boson or fermion canonical commutation relations,
\begin{align}
[\hat{\psi}_m(\bm{r}), \hat{\psi}^{\dagger}_{m'}(\bm{r}')]_{\sigma}&\equiv \hat{\psi}_m(\bm{r})\hat{\psi}_{m'}^{\dagger}(\bm{r}')-\sigma\hat{\psi}^{\dagger}_{m'}(\bm{r}')\hat{\psi}_m(\bm{r})\notag \\
&=\delta_{m,m'}\delta(\bm{r}-\bm{r}'),\label{eq:definition_of_commutation_relation}\\
[\hat{\psi}_m(\bm{r}), \hat{\psi}_{m'}(\bm{r}')]_{\sigma}&=[\hat{\psi}^{\dagger}_m(\bm{r}), \hat{\psi}_{m'}^{\dagger}(\bm{r}')]_{\sigma}=0,\label{eq:definition_of_commutation_relation_for_zero}
\end{align}
where $\sigma=+1$ for bosons and $\sigma=-1$ for fermions. We define the local-density operator $\hat{n}(\bm{r})\equiv \sum_m\hat{\psi}^{\dagger}_m(\bm{r})\hat{\psi}_m(\bm{r})$, the local magnetization operator $\hat{F}_{\mu}(\bm{r})\equiv \sum_{m,n}(f_{\mu})_{m n}\hat{\psi}^{\dagger}_m(\bm{r})\hat{\psi}_n(\bm{r})$, and the local nematic-tensor operator $\hat{N}_{\mu\nu}(\bm{r})\equiv \sum_{m,n}(f_{\mu}f_{\nu}+f_{\nu}f_{\mu})_{m n}\hat{\psi}^{\dagger}_m(\bm{r})\hat{\psi}_n(\bm{r})/2$, where $(f_{\mu})_{m n}$ $(\mu=x,y,z)$ is the $(m,n)$ component of the spin-$f$ matrix. The local raising and lowering operators are defined by $\hat{F}_{\pm}(\bm{r})\equiv\sum_{m,n}(f_{\pm})_{m n}\hat{\psi}^{\dagger}_m(\bm{r})\hat{\psi}_n(\bm{r})$, where $(f_{\pm})_{m n}\equiv (f_x\pm if_y)_{mn}$.

The Hamiltonian for the system is given by
\begin{align}
\hat{H}&\equiv \hat{H}_{\rm Kin}+\hat{H}_{\rm LZ}+\hat{H}_{\rm QZ}+\hat{H}_{\rm Int},\label{eq:spin-1_Hamiltonian}\\
\hat{H}_{\rm Kin}&\equiv \int d\bm{r}\sum_m\hat{\psi}^{\dagger}_m(\bm{r})\left(-\frac{\hbar^2}{2M}\nabla^2\right)\hat{\psi}_m(\bm{r}),\label{eq:kinetic_term}\\
\hat{H}_{\rm LZ}&\equiv -p\hat{M}_z,\label{eq:linear_zeeman_term}\\
\hat{H}_{\rm QZ}&\equiv q\hat{N}_{z z},\label{eq:quadratic_zeeeman_term}
\end{align}
where $M$ is the atomic mass, $p$ and $q$ are the magnitudes of the linear and quadratic Zeeman energies, $\hat{M}_{\alpha}\equiv \int d\bm{r}\hat{F}_{\alpha}(\bm{r})$ $(\alpha=x,y,z,+,-)$ is the total magnetization operator, and $\hat{N}_{\alpha\beta}\equiv \int d\bm{r}\hat{N}_{\alpha\beta}(\bm{r})$ is the total nematic-tensor operator. The interaction Hamiltonian is defined by \cite{Kawaguchi2012}
\begin{align}
\hat{H}_{\rm Int}&\equiv\sum_{\mathcal{F,M}}\frac{g_{\mathcal{F}}}{2}\int d\bm{r}\hat{A}^{\dagger}_{\mathcal{FM}}(\bm{r})\hat{A}_{\mathcal{FM}}(\bm{r})\notag \\
&=\frac{1}{2}\int \hspace{-0.1em}d\bm{r}\hspace{-0.75em}\sum_{m,n,m',n'}\hspace{-0.65em}C^{mm'}_{nn'}\hat{\psi}_m^{\dagger}(\bm{r})\hat{\psi}^{\dagger}_{m'}(\bm{r})\hat{\psi}_{n'}(\bm{r})\hat{\psi}_n(\bm{r}),\label{eq:interaction_term_C_continuous}\\
\hat{A}_{\mathcal{FM}}(\bm{r})&\equiv \sum_{m,m'}\bracket{\mathcal{F, M}}{f,m; f,m'}\hat{\psi}_m(\bm{r})\hat{\psi}_{m'}(\bm{r}),\label{eq:definition_AFM}\\
C^{mm'}_{nn'}&\equiv \sum_{\mathcal{F,M}}g_{\mathcal{F}}\bracket{f,m;f,m'}{\mathcal{F,M}}\bracket{\mathcal{F,M}}{f,n;f,n'},\label{eq:interaction_coefficient}
\end{align}
where $\mathcal{F}=0,2,\cdots, 2\lfloor f\rfloor$ and $\mathcal{M}=-\mathcal{F}, -\mathcal{F}+1,\cdots, \mathcal{F}$ are the total spin and its $z$ component of two colliding particles, respectively, $\lfloor\cdot \rfloor$ is the floor function, $g_{\mathcal{F}}$ is the coupling constant for the spin-$\mathcal{F}$ channel \cite{note_coupling}, $\hat{A}_{\mathcal{FM}}(\bm{r})$ is the pair annihilation operator with the spin state $(\mathcal{F}, \mathcal{M})$ at position $\bm{r}$, and $\bracket{\mathcal{F, M}}{f,m; f,m'}$ is the Clebsch-Gordan coefficient. The total particle number, the total momentum, and the $z$ component of the total magnetization are the conserved quantities:
\begin{align}
[\hat{H}, \hat{N}]=0, \quad [\hat{H}, \hat{\bm{P}}]=\bm{0}, \quad [\hat{H}, \hat{M}_z]=0,\label{eq:commutators}
\end{align}
where $\hat{N}\equiv \int d\bm{r}\hat{n}(\bm{r})$ and $\hat{\bm{P}}\equiv (-i\hbar/2)\int d\bm{r}\sum_m[\hat{\psi}^{\dagger}_m(\bm{r})\nabla\hat{\psi}_m(\bm{r})-{\rm H.c.}]$ (${\rm H.c.}$ denotes the Hermitian conjugate).

We consider two Hilbert subspaces $\mathcal{H}_{\rm F}$ and $\mathcal{H}_{\rm SW}^{\bm{k}}$. $\mathcal{H}_{\rm F}$ is spanned by the eigenvectors of the Hamiltonian $\{\cket{\Psi_{\rm F}}\}$ that satisfy $\hat{N}\cket{\Psi_{\rm F}}=N\cket{\Psi_{\rm F}}$, $\hat{\bm{P}}\cket{\Psi_{\rm F}}=\bm{0}$, and $\hat{M}_z\cket{\Psi_{\rm F}}=f N\cket{\Psi_{\rm F}}$. In this state, all $N$ particles occupy the magnetic sublevel $m=f$, {\it i.e.}, the fully spin-polarized ferromagnetic state. $\mathcal{H}_{\rm SW}^{\bm{k}}$ is spanned by the eigenvectors of the Hamiltonian $\{|\Psi_{\rm SW}^{\bm{k}}\rangle\}$ that satisfy $\hat{N}|\Psi_{\rm SW}^{\bm{k}}\rangle=N|\Psi_{\rm SW}^{\bm{k}}\rangle$, $\hat{\bm{P}}|\Psi^{\bm{k}}_{\rm SW}\rangle=\hbar\bm{k}|\Psi^{\bm{k}}_{\rm SW}\rangle$, and $\hat{M_z}|\Psi_{\rm SW}^{\bm{k}}\rangle=(fN-1)|\Psi_{\rm SW}^{\bm{k}}\rangle$, where $\bm{k}$ is the wave number of a spin wave. In $\mathcal{H}_{\rm SW}^{\bm{k}}$, $N-1$ particles occupy the magnetic sublevel $m=f$, and one particle occupies $m=f-1$.

We assume that the ground state $\cket{\rm g.s.}$ is the fully ferromagnetic state, {\it i.e.}, $\cket{\rm g.s.}$ is the lowest-energy eigenstate in $\mathcal{H}_{\rm F}$, where $\cket{\rm g.s.}$ is normalized as $\bracket{\rm g.s.}{\rm g.s.}=1$. The ground-state energy is denoted by $E_0$. We define the one-magnon excited state $\cket{\rm e. s.}$ as the lowest-energy eigenstate in $\mathcal{H}_{\rm SW}^{\bm{k}}$, where $\cket{\rm e.s.}$ is normalized as $\bracket{\rm e.s.}{\rm e.s.}=1$. The eigenenergy of the one-magnon excited state is denoted by $\epsilon_{\bm{k}}^{\rm SW}+E_0$, where $\epsilon_{\bm{k}}^{\rm SW}\ge 0$. A similar definition of the spin-wave excitation is used in Refs.~\cite{Tasaki1994,Tasaki1996}.

\subsection{Magnon excitation at the nean-field level}\label{sebsec:Magnon_mean_field}
Before presenting our results, we show that the magnon excitation energy for spin-$f$ bosons is the same as the dispersion of a free particle in the mean-field approximation. For simplicity, we consider the case of $p=q=0$. The Heisenberg equation for the field operator is given by
\begin{align}
i\hbar\frac{\partial}{\partial t}\hat{\psi}_m(\bm{r}, t)&=-\frac{\hbar^2}{2M}\nabla^2\hat{\psi}_m(\bm{r}, t)\notag \\
&\hspace{-2.0em}+\sum_{m',n,n'}C_{n n'}^{m m'}\hat{\psi}^{\dagger}_{m'}(\bm{r}, t)\hat{\psi}_{n'}(\bm{r}, t)\hat{\psi}_n(\bm{r}, t).\label{eq:Heisenberg_equation_for_hat_psi_complete_ver}
\end{align}
In the mean-field approximation \cite{Ohmi1998,Ho1998}, the system is described by the Gross-Pitaevskii equation, which is determined by replacing the field operator $\hat{\psi}_m(\bm{r}, t)$ in Eq.~(\ref{eq:Heisenberg_equation_for_hat_psi_complete_ver}) with the condensate wave function $\Psi_m(\bm{r}, t)$:
\begin{align}
i\hbar\frac{\partial}{\partial t}\Psi_m(\bm{r}, t)&=-\frac{\hbar^2}{2M}\nabla^2\Psi_m(\bm{r}, t)\notag \\
&\hspace{-2.0em}+\sum_{m',n,n'}C_{n n'}^{m m'}\Psi^{\ast}_{m'}(\bm{r}, t)\Psi_{n'}(\bm{r}, t)\Psi_n(\bm{r}, t).\label{eq:Gross-Pitaevskii_spin-f}
\end{align}
The ferromagnetic solution is given  by $\Psi_m(\bm{r}, t)=e^{-i\mu t/\hbar}\delta_{m,f}\sqrt{n_0}$, where $\mu=C^{f f}_{f f}n_0=g_{2f}n_0$ is the chemical potential and $n_0$ is the particle density. Substituting $e^{-i\mu t/\hbar}[\sqrt{n_0}\delta_{m,f}+\delta\Psi_m(\bm{r}, t)]$ into Eq.~(\ref{eq:Gross-Pitaevskii_spin-f}) and neglecting the higher-order terms of $\delta\Psi_m(\bm{r}, t)$, we obtain the equation of motion for the $m=f-1$ component:
\begin{align}
i\hbar\frac{\partial}{\partial t}\delta\Psi_{f-1}(\bm{r}, t)&=-\frac{\hbar^2}{2M}\nabla^2\delta\Psi_{f-1}(\bm{r}, t),\label{eq:equation_of_motion_for_magnon_mean_field}
\end{align}
where we used $C_{f f}^{f-1,f}=0$ and $C_{f,f-1}^{f-1,f}=C_{f-1,f}^{f,f-1}=g_{2f}/2$. This equation yields the magnon excitation energy $\epsilon_{\bm{k}}^{\rm MF}=\hbar^2\bm{k}^2/2M$, which is the same as the dispersion relation of a free particle.

\subsection{Result 1 (continuous systems)}\label{sebsec:Results1}

The one-magnon excitation energy $\epsilon_{\bm{k}}^{\rm SW}$ satisfies the following inequality:
\begin{align}
p-(2f-1)q\le \epsilon_{\bm{k}}^{\rm SW}\le\epsilon_{\bm{k}}^0+p-(2f-1)q,\label{eq:main_results_inequality}
\end{align}
where $\epsilon_{\bm{k}}^0\equiv \hbar^2\bm{k}^2/2M$ is the free-particle energy.

In the long-wavelength limit, we can expand $\epsilon_{\bm{k}}^{\rm SW}$ as $\epsilon_{\bm{k}}^{\rm SW}= p-(2f-1)q+\hbar^2\bm{k}^2/2M^{\ast}+O(|\bm{k}|^{3})$, where $M^{\ast}$ is the effective mass of the magnon. Therefore, Eq.~(\ref{eq:main_results_inequality}) gives the inequality
\begin{eqnarray}
M^{\ast}\ge M.\label{eq:inequality_effective_mass}
\end{eqnarray}
Thus, the effective mass never becomes lighter than the bare atomic mass. Since the upper bound in Eq.~(\ref{eq:main_results_inequality}) coincides with the excitation energy of the noninteracting systems, the contact interaction always enhances the effective mass of the magnon.

\subsection{Proof of result 1}\label{subsec:Proof_of_Results1}

Our strategy for proving Eq.~(\ref{eq:main_results_inequality}) is based on a variational method. We note that a similar method was used in Ref.~\cite{Yang2003} for two-component Bose systems with SU(2) symmetry. For $\cket{\Psi}\in\mathcal{H}_{\rm SW}^{\bm{k}}$, we can show that the following inequality holds:
\begin{align}
\frac{\bra{\Psi}\hat{H}\cket{\Psi}}{\bracket{\Psi}{\Psi}}\ge\epsilon_{\bm{k}}^{\rm SW}+E_0.\label{eq:inequality_for_variational_principle}
\end{align}
Here, we take $\cket{\Psi}\equiv \hat{F}_-(-\bm{k})\cket{\rm g.s.}$ as a trial wave function, where $F_-(\bm{k})\equiv \int d\bm{r}e^{-i\bm{k}\cdot\bm{r}}\hat{F}_-(\bm{r})=\hat{F}_+^{\dagger}(-\bm{k})$. One can easily show that $F_-(-\bm{k})\cket{\rm g.s.}\in\mathcal{H}_{\rm SW}^{\bm{k}}$. Using the relations $\hat{H}\cket{\rm g.s.}=E_0\cket{\rm g.s.}$ and $\hat{F}_+(\bm{k})\cket{\rm g.s.}=0$, Eq.~(\ref{eq:inequality_for_variational_principle}) is rewritten as
\begin{align}
\epsilon_{\bm{k}}^{\rm SW}\le \frac{\bra{\rm g.s.}[\hat{F}_+(\bm{k}), [\hat{H}, \hat{F}_-(-\bm{k})]]\cket{\rm g.s.}}{\bra{\rm g.s.}[\hat{F}_+(\bm{k}), \hat{F}_-(-\bm{k})]\cket{\rm g.s.}}.\label{eq:upper_bound_for_magnon_excitation}
\end{align}

To calculate the numerator of Eq.~(\ref{eq:upper_bound_for_magnon_excitation}), we first consider the case of $p=q=0$. In this case, the equation of continuity for the magnetization is given by \cite{note_equation_of_continuity}
\begin{align}
\frac{\partial}{\partial t}\hat{F}_{\mu}(\bm{r}, t)+\nabla\cdot\hat{\bm{J}}^{\rm spin}_{\mu}(\bm{r}, t)=0\label{eq:equation_of_continuity_continuous}
\end{align}
in the Heisenberg representation, where $\hat{\bm{J}}^{\rm spin}_{\mu}(\bm{r})\equiv -i\hbar/(2M)\sum_{m,n}[(f_{\mu})_{m n}\hat{\psi}^{\dagger}_m(\bm{r})\nabla\hat{\psi}_n(\bm{r})-{\rm H.c.}]$ is the spin-current operator. As in the case of the derivation of the $f$-sum rule \cite{Nozieres1966}, the equation of continuity in the $k$ space and the Heisenberg equation for $\hat{F}_{-}(\bm{k}, t)$ gives
\begin{align}
[\hat{H}, \hat{F}_-(-\bm{k})]=\hbar\bm{k}\cdot\hat{\bm{J}}^{\rm spin}_-(-\bm{k}),\label{eq:first_commutator}
\end{align}
where $\hat{\bm{J}}^{\rm spin}_-(\bm{k})\equiv \int d\bm{r}e^{-i\bm{k}\cdot\bm{r}}[\hat{\bm{J}}^{\rm spin}_x(\bm{r})-i\hat{\bm{J}}^{\rm spin}_y(\bm{r})]$. The detailed calculations of the interaction terms are shown in the Appendix \ref{app:calculation}. After some algebra, we obtain the numerator of Eq.~(\ref{eq:upper_bound_for_magnon_excitation}) as
\begin{align}
\bra{\rm g.s.}[\hat{F}_+(\bm{k}), [\hat{H}, \hat{F}_-(-\bm{k})]]\cket{\rm g.s.}=2fN\epsilon_{\bm{k}}^0,\label{eq:numerator_of_upper_bound}
\end{align}
where we used $M\int d\bm{r}\hat{\bm{J}}^{\rm spin}_z(\bm{r})\cket{\rm g.s.}=f\hat{\bm{P}}\cket{\rm g.s.}=\bm{0}$ and $\hat{N}_{z z}\cket{\rm g.s.}=f^2N\cket{\rm g.s.}$. The denominator of Eq.~(\ref{eq:upper_bound_for_magnon_excitation}) becomes
\begin{align}
\bra{\rm g.s.}[\hat{F}_+(\bm{k}), \hat{F}_-(-\bm{k})]\cket{\rm g.s.}=2f N.\label{eq:results_denominator_continuous}
\end{align}
We thus obtain the upper and lower bounds for $p=q=0$: $0\le\epsilon_{\bm{k}}^{\rm SW}\le \epsilon_{\bm{k}}^0$.

The case of $p\not=0$ and $q\not=0$ can be derived as follows. When $\hat{H}_{\rm LZ}$ and $\hat{H}_{\rm QZ}$ act on the state vector belonging to the Hilbert subspace $\mathcal{H}_{\rm F}$ or $\mathcal{H}_{\rm SW}^{\bm{k}}$, they reduce to $-p f N$ and $q f^2N$ for $\mathcal{H}_{\rm F}$ and $-p(fN-1)$ and $q(f^2N-2f+1)$ for $\mathcal{H}_{\rm SW}^{\bm{k}}$. Therefore, the linear and the quadratic Zeeman terms lift only the eigenenergy. The increase in the eigenenergy is given by $\bra{\rm e.s. }\hat{H}_{\rm LZ}+\hat{H}_{\rm QZ}\cket{\rm e.s.}-\bra{\rm g.s.}\hat{H}_{\rm LZ}+\hat{H}_{\rm QZ}\cket{\rm g.s.}=p-(2f-1)q$. Therefore, we obtain the inequality (\ref{eq:main_results_inequality}). The result in the presence of the uniform magnetic field is due to the fact that a magnon in the magnetic field is the massive Nambu-Goldstone mode \cite{Nicolis2013,Watanabe2013,Takahashi2015}.

\section{Lattice Systems}\label{sec:Lattice_systems}

We next consider the spin-$f$ bosons or fermions in $d$-dimensional lattice systems. The annihilation (creation) operator at the lattice site $\bm{R}$ is denoted by $\hat{a}_{\bm{R},m}$ $(\hat{a}_{\bm{R},m}^{\dagger})$, which satisfies the commutation relations $[\hat{a}_{\bm{R},m}, \hat{a}^{\dagger}_{\bm{R}',m'}]_{\sigma}=\delta_{\bm{R},\bm{R}'}\delta_{m,m'}$ and $[\hat{a}_{\bm{R},m}, \hat{a}_{\bm{R}',m'}]_{\sigma}=[\hat{a}^{\dagger}_{\bm{R},m}, \hat{a}_{\bm{R}',m'}^{\dagger}]_{\sigma}=0$. We define the local density operator $\hat{n}(\bm{R})\equiv \sum_m\hat{a}^{\dagger}_{\bm{R},m}\hat{a}_{\bm{R},m}$, the local magnetization operator $\hat{F}_{\alpha}(\bm{R})\equiv \sum_{m,n}(f_{\alpha})_{m n}\hat{a}^{\dagger}_{\bm{R},m}\hat{a}_{\bm{R},n}$, and the local nematic-tensor operator $\hat{N}_{\alpha\beta}(\bm{R})\equiv \sum_{m,n}(f_{\alpha}f_{\beta}+f_{\beta}f_{\alpha})_{m n}\hat{a}^{\dagger}_{\bm{R},m}\hat{a}_{\bm{R},n}/2$. We also introduce the annihilation (creation) operator of the Bloch state $\hat{b}_{\bm{k}, m}$ $(\hat{b}^{\dagger}_{\bm{k},m })$, where $\hbar\bm{k}$ is the crystal momentum. The relation between $\hat{a}_{\bm{R},m}$ and $\hat{b}_{\bm{k},m}$ is given by $\hat{a}_{\bm{R},m}=(1/\sqrt{N_{\rm L}})\sum_{\bm{k}}e^{i\bm{k}\cdot\bm{R}}\hat{b}_{\bm{k},m}$, where $N_{\rm L}$ is the number of sites.

The single-band Hubbard Hamiltonian of the system is given by
\begin{align}
\hat{H}&\equiv \hat{H}_{\rm Kin}+\hat{H}_{\rm LZ}+\hat{H}_{\rm QZ}+\hat{H}_{\rm Int},\label{eq:Hamiltonian_for_lattice_system}\\
\hat{H}_{\rm Kin}&\equiv \sum_{\bm{R},\bm{R}',m}t_{\bm{R},\bm{R}'}\hat{a}^{\dagger}_{\bm{R},m}\hat{a}_{\bm{R}',m}=\sum_{\bm{k},m}\epsilon(\bm{k})\hat{b}_{\bm{k},m}^{\dagger}\hat{b}_{\bm{k},m},\label{eq:Kinetic_Hamiltonian_lattice}\\
\hat{H}_{\rm LZ}&\equiv -p\hat{M}_z,\label{eq:Linear_Zeeman_Hamiltonian_lattice}\\
\hat{H}_{\rm QZ}&\equiv q\hat{N}_{z z},\label{eq:Quadratic_Zeeman_Hamiltonian_lattice}
\end{align}
where $t_{\bm{R},\bm{R}'}=t_{\bm{R}',\bm{R}}\in\mathbb{R}$ is the hopping parameter depending only on the distance $|\bm{R}-\bm{R}'|$, $\epsilon(\bm{k})$ is the bare band dispersion, $\hat{M}_{\alpha}\equiv \sum_{\bm{R}}\hat{F}_{\alpha}(\bm{R})$ is the total magnetization operator, and $\hat{N}_{\alpha\beta}\equiv \sum_{\bm{R}}\hat{N}_{\alpha\beta}(\bm{R})$ is the total nematic-tensor operator. The interaction Hamiltonian is defined by
\begin{align}
\hat{H}_{\rm Int}&\equiv \sum_{\mathcal{F,M},\bm{R}}\frac{\tilde{g}_{\mathcal{F}}}{2}\hat{A}^{\dagger}_{\mathcal{FM}}(\bm{R})\hat{A}_{\mathcal{FM}}(\bm{R}),\label{eq:Interaction_Hamiltonian_lattice}
\end{align}
where $\tilde{g}_{\mathcal{F}}$ is the interaction strength of the spin-$\mathcal{F}$ channel and $\hat{A}_{\mathcal{FM}}(\bm{R})\equiv \sum_{m,m'}\bracket{\mathcal{F, M}}{f,m;f,m'}\hat{a}_{\bm{R},m}\hat{a}_{\bm{R},m'}$ is the pair annihilation operator of the spin state $(\mathcal{F}, \mathcal{M})$ at site $\bm{R}$. We note that the total particle number operator $\hat{N}\equiv \sum_{\bm{R}}\hat{n}(\bm{R})$, the $z$ component of the total magnetization operator $\hat{M}_z$, and the translation operator $\hat{T}_{\bm{R}}$ commute with $\hat{H}$, where $\hat{T}_{\bm{R}}$ satisfies the relation $\hat{T}_{\bm{R}}\hat{a}_{\bm{R}',m}\hat{T}^{-1}_{\bm{R}}=\hat{a}_{\bm{R}+\bm{R}',m}$.

In the same manner as in the continuous systems, we use the two Hilbert subspaces $\mathcal{H}_{\rm F}$ and $\mathcal{H}_{\rm SW}^{\bm{k}}$. $\mathcal{H}_{\rm F}$ and $\mathcal{H}_{\rm SW}^{\bm{k}}$ are spanned by the eigenvectors of the Hamiltonian $\cket{\Psi_{\rm F}}$ and $|\Psi_{\rm SW}^{\bm{k}}\rangle$ that satisfy $\hat{N}\cket{\Psi_{\rm F}}=N\cket{\Psi_{\rm F}}$, $\hat{M}_z\cket{\Psi_{\rm F}}=fN\cket{\Psi_{\rm F}}$, and $\hat{T}_{\bm{R}}\cket{\Psi_{\rm F}}=\cket{\Psi_{\rm F}}$ and $\hat{N}|\Psi_{\rm SW}^{\bm{k}}\rangle=N|\Psi_{\rm SW}^{\bm{k}}\rangle$, $\hat{M}_z|\Psi_{\rm SW}^{\bm{k}}\rangle=(fN-1)|\Psi_{\rm SW}^{\bm{k}}\rangle$, and $\hat{T}_{\bm{R}}|\Psi_{\rm SW}^{\bm{k}}\rangle=e^{-i\bm{k}\cdot\bm{R}}|\Psi_{\rm SW}^{\bm{k}}\rangle$, respectively. We assume that the ground state $\cket{\rm g.s.}$ is the lowest-energy eigenstate in $\mathcal{H}_{\rm F}$, normalized as $\bracket{\rm g.s.}{\rm g.s.}=1$, and its energy is $E_0$. We also assume that the one-magnon excited state is the lowest-energy eigenstate in $\mathcal{H}_{\rm SW}^{\bm{k}}$ with an energy $\epsilon_{\bm{k}}^{\rm SW}+E_0$.

\subsection{Result 2 (lattice systems)}\label{subsec:Result_2_lattice_system}

The one-magnon excitation energy satisfies
\begin{align}
&p-(2f-1)q\le \epsilon_{\bm{k}}^{\rm SW}\le \tilde{\epsilon}(\bm{k})+p-(2f-1)q,\label{eq:Results_2_lattice}
\end{align}
where
\begin{align}
\tilde{\epsilon}(\bm{k})\equiv \frac{1}{N}\sum_{\bm{k}'}[\epsilon(\bm{k}'+\bm{k})-\epsilon(\bm{k}')]\bra{\rm g.s.}\hat{n}_{\bm{k}',f}\cket{\rm g.s.},\label{eq:definition_of_epsilon_tilde_lattice}
\end{align}
with $\hat{n}_{\bm{k},m}\equiv \hat{b}^{\dagger}_{\bm{k},m}\hat{b}_{\bm{k},m}$. In the long-wavelength limit, $\tilde{\epsilon}(\bm{k})$ can be expanded as
\begin{align}
\tilde{\epsilon}(\bm{k})&\simeq \sum_{i,j}\frac{\hbar^2k_ik_j}{2N}\sum_{\bm{k}'}\frac{\bra{\rm g.s.}\hat{n}_{\bm{k}',f}\cket{\rm g.s.}}{m^{\ast}_{ij}(\bm{k}')},\label{eq:epsilon_tilde_long_wavelength_lattice}\\
[m_{i j}^{\ast}(\bm{k})]^{-1}&\equiv \frac{1}{\hbar^2}\frac{\partial^2}{\partial k_i\partial k_j}\epsilon(\bm{k}),\label{eq:definition_of_effective_mass_lattice_systems}
\end{align}
where we assume the space-inversion symmetry of the system and $m_{i j}^{\ast}(\bm{k})$ is an effective mass tensor.

The upper bound in Eq.~(\ref{eq:Results_2_lattice}) reflects the bare band dispersion, in contrast to the uniform continuous systems. This is due to the lack of the continuous translational symmetry of the system. We note that a similar situation appears in the $f$-sum rule for lattice systems \cite{Huber2007}.

\subsection{Proof of result 2}\label{subsec:Proof_of_result_2}

The proof is almost the same as that for the continuous systems. In Eq.~(\ref{eq:inequality_for_variational_principle}), we take $\cket{\Psi}\equiv \hat{F}_-(-\bm{k})\cket{\rm g.s.}$, where $\hat{F}_-(\bm{k})\equiv \sum_{\bm{R}} e^{-i\bm{k}\cdot\bm{R}}\hat{F}_{-}(\bm{R})$. The upper bound of the lattice systems is given by the same expression as the continuous systems (\ref{eq:upper_bound_for_magnon_excitation}).

As in the case of the continuous systems, we first consider the zero magnetic field $p=q=0$. The commutator $[\hat{H}, \hat{F}_-(-\bm{k})]$ and its double commutator are calculated to be
\begin{align}
&[\hat{H}, \hat{F}_-(-\bm{k})]\notag \\
&=\sum_{\bm{k}', m,n}(f_{-})_{m n}\left[\epsilon(\bm{k}'+\bm{k})-\epsilon(\bm{k}')\right]\hat{b}^{\dagger}_{\bm{k}'+\bm{k},m}\hat{b}_{\bm{k}',n},\label{eq:first_commutator_lattice}\\
&[\hat{F}_+(\bm{k}), [\hat{H}, \hat{F}_-(-\bm{k})]]\notag \\
&=f(f+1)\sum_{\bm{k}',m}[\epsilon(\bm{k}'-\bm{k})+\epsilon(\bm{k}'+\bm{k})-2\epsilon(\bm{k}')]\hat{n}_{\bm{k}',m}\notag \\
&-\sum_{\bm{k}',m,n}(f_z^2)_{m n}[\epsilon(\bm{k}'-\bm{k})+\epsilon(\bm{k}'+\bm{k})-2\epsilon(\bm{k}')]\hat{b}^{\dagger}_{\bm{k}',m}\hat{b}_{\bm{k}',n}\notag \\
&+\sum_{\bm{k}',m,n}(f_z)_{m n}[\epsilon(\bm{k}'+\bm{k})-\epsilon(\bm{k}'-\bm{k})]\hat{b}^{\dagger}_{\bm{k}',m}\hat{b}_{\bm{k}',n}.\label{eq:double_commutator_H_lattice}
\end{align}
Using the expectation values for the ground state, $\bra{\rm g.s.}\hat{n}_{\bm{k},m}\cket{\rm g.s.}=\delta_{m,f}\bra{\rm GS}\hat{n}_{\bm{k},f}\cket{\rm g.s.}$ and $\bra{\rm g.s.}(f_z^s)_{m n}\hat{b}^{\dagger}_{\bm{k},m}\hat{b}_{\bm{k},n}\cket{\rm g.s.}=f^s\delta_{m,f}\delta_{n,f}\bra{\rm g.s.}\hat{b}^{\dagger}_{\bm{k},f}\hat{b}_{\bm{k},f}\cket{\rm g.s.}$ ($s=1,2$), we obtain
\begin{align}
\bra{\rm g.s.}[\hat{F}_+(\bm{k}), [\hat{H}, \hat{F}_-(-\bm{k})]]\cket{\rm g.s.}=2fN\tilde{\epsilon}(\bm{k}).\label{eq:expectation_value_of_numerator_lattice}
\end{align}
The denominator in Eq.~(\ref{eq:upper_bound_for_magnon_excitation}) is $\bra{\rm GS}[\hat{F}_+(\bm{k}), \hat{F}_-(-\bm{k})]\cket{\rm GS}=2fN$, which is the same as that of the continuous systems.

The upper and lower bounds for lattice systems at the zero magnetic field thus become $0\le \epsilon_{\bm{k}}^{\rm SW}\le \tilde{\epsilon}(\bm{k})$. In a manner similar to that in the continuous systems, the magnetic field lifts only the eigenenergy. Therefore, we obtain inequality (\ref{eq:Results_2_lattice}).

\section{Concluding Remarks}\label{sec:Concluding_Remarks}

We have derived the exact upper bound of the one-magnon excitation energy for the ferromagnetic spinor Bose and Fermi gases in both continuous and lattice systems, respectively. This result holds for arbitrary spin and spatial dimensions under the following assumptions: (i) the ground state is fully spin polarized and has zero  (crystal) momentum, (ii) the Hamiltonian has continuous (discrete) translational symmetry in continuous (lattice) systems, and (iii) two-body interaction is the $s$-wave contact one and is isotropic in spin-space. Since our results are based on the local spin-conservation law, analogous to the $f$-sum rule due to the particle number conservation \cite{Nozieres1966}, the inequality (\ref{eq:main_results_inequality}) is one of the universal properties of the ferromagnetic spinor gases. Inequalities (\ref{eq:main_results_inequality}) and (\ref{eq:inequality_effective_mass}) serve as a check for the validity of quantum many-body corrections. We note that the quantum many-body correction to the magnon mass recently calculated in Ref.~\cite{Phuc2013_2} satisfies inequality (\ref{eq:main_results_inequality}).

From inequality (\ref{eq:main_results_inequality}), we can conclude that the effective mass of a magnon never becomes lighter than the atomic mass. Since the equality $M^{\ast}=M$ holds for non-interacting systems, we find that the contact interactions always increase the effective mass of a magnon. If one observes the effective mass of a magnon that is lighter than the atomic mass, its origin is not the isotropic contact interaction but other effects, such as the magnetic dipole-dipole interaction, which is long ranged and anisotropic. In fact, we showed that the effective mass of a magnon can decrease due to the magnetic dipole-dipole interaction \cite{Saito2015_a}.

Our results are valid not only for ferromagnetic spinor Bose gases but also for ferromagnetic Fermi gases. In contrast to the spinor Bose gases (see Ref.~\cite{Katsura2013}), the ferromagnetism in Fermi systems is a generally difficult problem since the ferromagnetic phase is realized in a strongly correlated regime. The Nagaoka state is an example of a fully spin-polarized ground state in the spin-$1/2$ Hubbard model \cite{Nagaoka1965,Tasaki1989} and SU($N)$ Hubbard model \cite{Katsura_Tanaka_2013}. For two-component $(f=1/2)$ Fermi gases, the Stoner instability does not occur due to the competition with the pairing instability \cite{Sanner2012}. Fermi gases with higher spin have already been realized in the experiments \cite{Taie2010,Krauser2012}. The ferromagnetic phase transition in the SU($N$) Fermi systems has been theoretically discussed \cite{Cazalilla2009}.

Future work is to improve the upper and lower bounds of the excitation energy. Our upper bound (\ref{eq:main_results_inequality}) does not depend on the interaction strength. An appropriate modification of the trial wave function allows us to incorporate the effect of interactions into the upper and lower bounds, which may corroborate the Beliaev theory \cite{Phuc2013_2}. Another prospect will be the upper and lower bounds for the spin-wave excitations in other magnetic phases such as polar and antiferromagnetic phases.

\begin{acknowledgments}
We thank Hosho Katsura, Shintaro Hoshino, and Shinji Koshida for useful comments. This work was supported by MEXT KAKENHI Grant Number 25103007 and JSPS KAKENHI Grant Number 26400414.
\end{acknowledgments}


\appendix
\section{Calculation for Commutators}\label{app:calculation}
\begin{widetext}
In this appendix, we present the detailed calculations for the commutation relation $[\hat{H}_{\rm Int}, \hat{F}_{\mu}(\bm{r})] \;(\mu=x,y,z)$. According to Ref.~\cite{Kudo2011}, the interaction coefficient $C_{n n'}^{m m'}$ can be written as
\begin{align}
C_{nn'}^{mm'}&=\Lambda_0\delta_{m,n}\delta_{m',n'}+\sum_{k=1}^{\lfloor f \rfloor}\sum_{\nu_1,\cdots,\nu_k}\Lambda_k(f_{\nu_1}\cdots f_{\nu_k})_{mn}(f_{\nu_1}\cdots f_{\nu_k})_{m'n'},\label{eq:app_interaction_coefficients_other_representation}
\end{align}
where $\Lambda_0$ and $\Lambda_k$ are the linear combination of $g_{\mathcal{F}}$. The interaction Hamiltonian (\ref{eq:interaction_term_C_continuous}) reduces to
\begin{align}
\hat{H}_{\rm Int}&=\frac{\Lambda_0}{2}\int d\bm{r}:\hat{n}(\bm{r})^2:+\sum_{k=1}^{\lfloor f\rfloor}\frac{\Lambda_k}{2}\int d\bm{r}\sum_{\nu_1,\cdots,\nu_k}:\hat{\mathcal{N}}^{(k)}_{\nu_1\nu_2,\cdots,\nu_k}(\bm{r})^2:\label{eq:app_interaction_Hamiltonian_tensor_form}\\
&\equiv \hat{H}_{\rm Int}^{(0)}+\sum_{k=1}^{\lfloor f\rfloor}\hat{H}_{\rm Int}^{(k)},\label{eq:app_definition_H_int^k}\\
\hat{\mathcal{N}}^{(k)}_{\nu_1\cdots\nu_k}(\bm{r})&\equiv \sum_{m,n}(f_{\nu_1}\cdots f_{\nu_k})_{mn}\hat{\psi}^{\dagger}_m(\bm{r})\hat{\psi}_n(\bm{r}),\label{eq:app_definitino_of_non-symmetrized_nematic_tensor}
\end{align}
where $:\;\;:$ denotes the normal ordering and $\hat{\mathcal{N}}^{(k)}_{\nu_1\cdots\nu_k}(\bm{r})$ is the rank-$k$ local nematic-tensor operator. First, we calculate $[\hat{H}_{\rm Int}^{(0)},\hat{F}_{\mu}(\bm{r})]$:
\begin{align}
[\hat{H}_{\rm Int}^{(0)},\hat{F}_{\mu}(\bm{r})]&=\frac{\Lambda_0}{2}\sum_{m,n,m',n'}\int d\bm{r}'(f_{\mu})_{m'n'}[\hat{\psi}^{\dagger}_m(\bm{r}')\hat{\psi}^{\dagger}_n(\bm{r}')\hat{\psi}_n(\bm{r}')\hat{\psi}_m(\bm{r}'), \hat{\psi}_{m'}^{\dagger}(\bm{r})\hat{\psi}_{n'}(\bm{r})]\notag \\
&=\frac{\Lambda_0}{2}\sum_{m,n,m',n'}(f_{\mu})_{m'n'}\left[-\sigma\delta_{n,n'}\hat{\psi}^{\dagger}_{m'}(\bm{r})\hat{\psi}^{\dagger}_m(\bm{r})\hat{\psi}_n(\bm{r})\hat{\psi}_m(\bm{r})-\delta_{m,n'}\hat{\psi}^{\dagger}_{m'}(\bm{r})\hat{\psi}^{\dagger}_n(\bm{r})\hat{\psi}_n(\bm{r})\hat{\psi}_m(\bm{r})\right.\notag \\
&\left.\hspace{10.75em}+\delta_{m,m'}\hat{\psi}^{\dagger}_m(\bm{r})\hat{\psi}_n^{\dagger}(\bm{r})\hat{\psi}_n(\bm{r})\hat{\psi}_{n'}(\bm{r})+\sigma \delta_{n,m'}\hat{\psi}^{\dagger}_m(\bm{r})\hat{\psi}^{\dagger}_{n}(\bm{r})\hat{\psi}_m(\bm{r})\hat{\psi}_{n'}(\bm{r})\right]\notag \\
&=0.\label{eq:app_result_commutation_relation_Hint_and_Fmu}
\end{align}
Next, we calculate $[\hat{H}_{\rm Int}^{(k)}, \hat{F}_{\mu}(\bm{r})]$. We use a short-hand notation,
\begin{eqnarray}
(T_k)_{mn}\equiv (f_{\nu_1}\cdots f_{\nu_k})_{mn}.\label{eq:app_definition_T_k_mn}
\end{eqnarray}
The commutator $[\hat{H}_{\rm Int}^{(k)},\hat{F}_{\mu}(\bm{r})]$ is calculated to be
\begin{align}
[\hat{H}_{\rm Int}^{(k)}, \hat{F}_{\mu}(\bm{r})]&=\frac{\Lambda_k}{2}\int d\bm{r}'\sum_{m,n,m',n',m'',n''}\sum_{\nu_1,\cdots,\nu_k}(T_k)_{m n}(T_k)_{m' n'}(f_{\mu})_{m''n''}\notag \\
&\hspace{8.0em}\times[\hat{\psi}^{\dagger}_{m}(\bm{r}')\hat{\psi}^{\dagger}_{m'}(\bm{r}')\hat{\psi}_{n'}(\bm{r}')\hat{\psi}_{n}(\bm{r}'), \hat{\psi}^{\dagger}_{m''}(\bm{r})\hat{\psi}_{n''}(\bm{r})]\notag \\
&=\frac{\Lambda_k}{2}\sum_{m,n,m',n',m'',n''}\sum_{\nu_1,\cdots,\nu_k}(T_k)_{m n}(T_k)_{m' n'}(f_{\mu})_{m''n''}\notag \\
&\hspace{8.0em}\times\left[-\sigma\delta_{m',n''}\hat{\psi}^{\dagger}_{m''}(\bm{r})\hat{\psi}^{\dagger}_m(\bm{r})\hat{\psi}_{n'}(\bm{r})\hat{\psi}_{n}(\bm{r})-\delta_{m,n''}\hat{\psi}^{\dagger}_{m''}(\bm{r})\hat{\psi}^{\dagger}_{m'}(\bm{r})\hat{\psi}_{n'}(\bm{r})\hat{\psi}_{n}(\bm{r})\right.\notag \\
&\left.\hspace{10.0em}+\delta_{n,m''}\hat{\psi}^{\dagger}_m(\bm{r})\hat{\psi}^{\dagger}_{m'}(\bm{r})\hat{\psi}_{n'}(\bm{r})\hat{\psi}_{n''}(\bm{r})+\sigma\delta_{n',m''}\hat{\psi}^{\dagger}_m(\bm{r})\hat{\psi}^{\dagger}_{m'}(\bm{r})\hat{\psi}_{n}(\bm{r})\hat{\psi}_{n''}(\bm{r})\right]\notag \\
&=\Lambda_k\sum_{m,n,m',n'}\sum_{\nu_1,\cdots,\nu_k}(T_k)_{mn}[T_k, f_{\mu}]_{m'n'}\hat{\psi}^{\dagger}_m(\bm{r})\hat{\psi}^{\dagger}_{m'}(\bm{r})\hat{\psi}_{n'}(\bm{r})\hat{\psi}_{n}(\bm{r}).\label{eq:app_commutation_relation_Hint_k_and_Fmu_totyu}
\end{align}
Here, the commutator $[T_k, f_{\mu}]_{m n}$ becomes
\begin{align}
[T_k, f_{\mu}]_{mn}&=[f_{\nu_1}f_{\nu_2}\cdots f_{\nu_k}, f_{\mu}]_{m n}\notag \\
&=\sum_{j=1}^k \left(f_{\nu_1}\cdots f_{\nu_{j-1}}[f_{\nu_j}, f_{\mu}]f_{\nu_{j+1}}\cdots f_{\nu_k}\right)_{m n}\notag \\
&=\sum_{j=1}^k\sum_{\lambda}i\epsilon_{\nu_j \mu \lambda}\left(f_{\nu_1}\cdots f_{\nu_{j-1}}f_{\lambda}f_{\nu_{j+1}}\cdots f_{\nu_k}\right)_{m n},\label{eq:app_commutator_Tk_Fmu}
\end{align}
where $\epsilon_{\mu\nu\lambda}$ is the completely antisymmetric tensor. From Eqs.~(\ref{eq:app_commutation_relation_Hint_k_and_Fmu_totyu}) and (\ref{eq:app_commutator_Tk_Fmu}), we obtain
\begin{align}
[\hat{H}_{\rm Int}^{(k)}, \hat{F}_{\mu}(\bm{r})]&=\Lambda_k\sum_{m,n,m',n'}\sum_{\nu_1,\cdots,\nu_k,\lambda}\sum_{j=1}^k i\epsilon_{\nu_j\mu\lambda}(f_{\nu_1}\cdots f_{\nu_j}\cdots f_{\nu_k})_{mn}(f_{\nu_1}\cdots f_{\lambda}\cdots f_{\nu_k})_{m'n'}\notag \\
&\hspace{13.0em}\times \hat{\psi}^{\dagger}_m(\bm{r})\hat{\psi}^{\dagger}_{m'}(\bm{r})\hat{\psi}_{n'}(\bm{r})\hat{\psi}_{n}(\bm{r})\notag \\
&=-\Lambda_k\sum_{m,n,m',n'}\sum_{\nu_1,\cdots,\nu_k,\lambda}\sum_{j=1}^k i\epsilon_{\nu_j\mu\lambda}(f_{\nu_1}\cdots f_{\nu_j}\cdots f_{\nu_k})_{mn}(f_{\nu_1}\cdots f_{\lambda}\cdots f_{\nu_k})_{m'n'}\notag \\
&\hspace{13.0em}\times \hat{\psi}^{\dagger}_{m}(\bm{r})\hat{\psi}^{\dagger}_{m'}(\bm{r})\hat{\psi}_{n'}(\bm{r})\hat{\psi}_{n}(\bm{r})\notag \\
&=-[\hat{H}^{(k)}_{\rm Int}, \hat{F}_{\mu}(\bm{r})]\notag \\
&=0.\label{eq:app_commutator_Hint_k_and_Fmu}
\end{align}

Therefore, the commutation relation $[\hat{H}_{\rm Int}, \hat{F}_-(\bm{r})]=0$ holds for spin-$f$ bosons and fermions with the contact interaction. In the same manner, we can show $[\hat{H}_{\rm Int}, \hat{F}_{\mu}(\bm{R})]=0$ for the lattice system.

\end{widetext}


\begin{thebibliography}{99}
\bibitem{note_coupling}
For three-dimensional systems, the expression of the coupling constant is given by $g_{\mathcal{F}}=4\pi\hbar^2a_{\mathcal{F}}/M$, where $a_{\mathcal{F}}$ is the $s$-wave scattering length of the spin-$\mathcal{F}$ channel.
\bibitem{note_equation_of_continuity}
In general, this equation of continuity holds for the contact interaction. For nonlocal interactions, nonlocal terms can arise in the equation of continuity.
\end{thebibliography}

\end{document}